\documentstyle[aps,prb,multicol]{revtex}
\input{epsf.sty}
\begin{document}
\draft
\title{
Superconducting Fluctuation Effects on the Electron Spin Susceptibility in YBa$_2$Cu$_3$O$_{6.95}$
}
\author{
   H. N. Bachman, V. F. Mitrovi{\'c}, A. P. Reyes\cite{reyesaddr}, 
W. P. Halperin, M. Eschrig, J. A. Sauls
}
\address{
   Department of Physics and Astronomy, and
   Science and Technology Center for Superconductivity,\\
   Northwestern University, Evanston, Illinois 60208
}
\author{
   A. Kleinhammes\cite{kleinhaddr}, P. Kuhns, W. G. Moulton
}
\address{
   National High Magnetic Field Laboratory
   Tallahassee, Florida 32310
}
\date{Received 30 July 1998}
\maketitle
\begin{abstract}
   The electronic spin susceptibility of YBa$_2$Cu$_3$O$_{6.95}$ has been measured with high precision up to 24 Tesla with 
   $^{17}$O nuclear magnetic resonance. Its temperature dependence can be accounted for by 
   superconducting fluctuations that result in a smooth crossover from the normal to the vortex liquid state. 
A magnetic field-temperature phase diagram for this crossover has been established having strong upward curvature.
\end{abstract}
\vspace{-11pt}
\begin{multicols}{2}
   The upper critical field is large in high-$T_c$ materials because of their small superconducting coherence lengths. However, a precise
determination of
$H_{c2}(T)$ at high magnetic fields is difficult because there is no well-defined signature of a phase transition. In magnetic fields
significantly larger  than the lower critical field
($H_{c1}(0) \approx$ 100 G) the transition 
   is broadened by the opening of a pseudogap in the electronic excitation spectrum. We performed a detailed
experimental and theoretical study of the fluctuation effects on the spin susceptibility in optimally doped YBCO and
developed a quantitative understanding for the onset of superconductivity solely in terms of a pairing pseudogap. This
allows us to define a crossover field, 
$H_{c2}(T)$, up to high magnetic fields.

				In high-$T_c$ 
   materials the effects of a magnetic field on pairing fluctuations have been discussed for the
diamagnetic response,\cite{welp89,lee89}  resistivity,\cite{kwok94} and 
   heat capacity.\cite{roulin96}  In general, efforts to determine $H_{c2}(T)$ have relied on {\it ad hoc} criteria that are not 
   related to superconducting fluctuations. The specific heat has been analyzed by Roulin {\it et al.}\cite{roulin96} using several criteria 
to determine $H_{c2}(T)$.
   SQUID-based magnetization measurements of Welp {\it et al.}\cite{welp89} show rounding in $M(T)$ near the 
   expected transition temperature as the field is increased. In this case the transition temperature was 
   determined by linear extrapolation of the temperature dependent diamagnetism in the superconducting state to intersect 
   with the normal state magnetization.

In the present work we determine $H_{c2}(T)$ from measurements of
   the Pauli spin susceptibility, $\chi_s$. We measure the $^{17}$O nuclear magnetic resonance (NMR) Knight shift
   and isolate the contribution of $\chi_s$, 
   taking into account orbital, diamagnetic, and vortex shifts. Because of the high precision of $^{17}$O NMR, we 
   measure the temperature dependence of $\chi_s$ to better than 0.1\% of the total normal state $\chi_s$ value. Thus, we are 
   able to make a quantitative comparison of the data with the theory of pairing fluctuation
corrections to $\chi_s$ 
in the normal state. In the superconducting state there is a mean-field region where 
$\chi_s$ is linear in temperature. A crossing-point analysis defines
   the crossover temperature, $T_c(H)$. 
We performed this analysis in magnetic fields up to 24 T.

Our aligned powder sample is the same as that studied previously\cite{reyes97a,takigawa89} having 30-40\% $^{17}$O-enriched 
YBa$_2$Cu$_3$O$_{6.95}$ prepared 
   by solid state reaction. Low field magnetization data show a sharp $T_c$ at 92.5 K.  The 
   $^{17}$O NMR spectra were obtained from the fast Fourier transform of a Hahn echo sequence: $\pi /2$-$\tau$-$\pi $-acquire and 
   only the $(1/2 \leftrightarrow -1/2)$ transition of the O(2,3) sites was studied.  High RF power allowed the use of short ($\approx 1.5 $
   $\mu$s) $\pi/2 $ pulses (2.5 $\mu$s at 2.1 T), giving a useful bandwidth $>100$ kHz.  
The $^{17}$O(2,3) \mbox{($1/2 \leftrightarrow -1/2$)} resonance has a low frequency tail owing to
oxygen deficiency in a small portion of the sample.\cite{reyes97a} Its effect on our
measurements was eliminated by performing a nonlinear least-squares fit
in the frequency domain which isolates the dominant, narrow spectrum of
optimally doped YBCO.
Temperature stability was $\pm \, 0.1$ K or 
   better.  For $H_o \le 14.9$ T we used superconducting magnets.  The shifts measured at 8.4 T were the most precise. The high field 
   measurements, 18.7 T to 24 T, were performed in a Bitter magnet adapted for modest homogeneity NMR at the 
   National High Magnetic Field Laboratory in Tallahassee, Florida.  High voltage arcing and temperature stability 
   in the constrained space of the Bitter magnet required that we design a special probe.\cite{reyes97b}

   Excluding temperature-independent quadrupolar terms, the frequency of $^{17}$O NMR in YBCO can be written,
\begin{eqnarray}
\label{eq1}
           \nu = ^{17}\! \! \gamma \left[ H_o(1+^{17}\!\!K_{\mbox{\small spin}} + ^{17}\!\!K_{\mbox{\small orb}}) + \Delta H_{\mbox{\small
dia}} +
\Delta H_{\mbox{\small v}}
\right] .  \nonumber
\end{eqnarray}
The gyromagnetic ratio of $^{17}$O is
$^{17}\gamma $; $H_o$ is the 
static magnetic field applied parallel to 
   the $\hat{c}$-axis; $\Delta H_{\mbox{\small v}}$ is a spatially varying field
that results from
pinned vortices, appearing below the melting or pinning temperature.\cite{reyes97a}
\begin{figure}[h]
\centerline{\epsfxsize0.90\hsize\epsffile{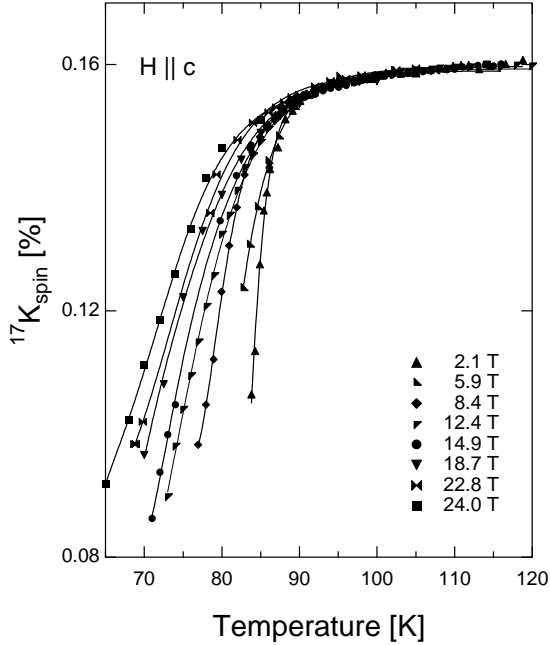}}
\begin{minipage}{0.95\hsize}
\caption[]{\label{KvTALL}\small
   The $^{17}$O(2,3) spin shift at different magnetic fields for YBCO$_{6.95}$. The spin shift is proportional to the Pauli spin susceptibility.
Lines are guides to the eye. }
\end{minipage}
\end{figure}
\noindent 
The orbital shift, $^{17}\!K_{\mbox{\small orb}}$, is
 small and temperature 
independent.  The diamagnetic 
   contributions, $\Delta H_{\mbox{\small dia}}$, whose 
origin is from surface supercurrents, are negligible at 
large fields.\cite{takigawa89}
In the vortex liquid state the inhomogeneous vortex fields, $\Delta H_{\mbox{\small v}}$, are 
   motionally averaged. 
At low temperatures vortices are pinned, and the resonance for $^{17}$O nuclei in their vicinity is shifted to
lower frequencies by
$^{17}\gamma\Delta H_{\mbox{\small v}}$ compared to nuclei in the vortex liquid 
phase. This shift allows us to identify the $^{17}$O resonance
associated with the vortex liquid
.\cite{reyes97a}  The temperature dependence of this part of the spectrum is given by
$^{17}\!K_{\mbox{\small spin}}$, which is 
   proportional to the electron spin susceptibility, $\chi_s$. The high sensitivity of $^{17}$O(2,3) NMR to $\chi_s$, via the
$\hat{c}$-axis hyperfine coupling, is an advantage compared with that of $^{63}$Cu(2) 
   where the $\hat{c}$-axis coupling is quite small and the copper resonance is broad.

   In \mbox{Fig. \ref{KvTALL}} we show the temperature-dependent shifts for $^{17}$O(2,3) measured over a
magnetic field range of 2.1 T to 24 T. 
  The normal state values at 120 K are fixed to the value of
$^{17}\!K_{\mbox{\small spin}}$ in the normal state, 
   0.16$\% \, \pm \, 0.01\,$\%. $^{17}\!K_{\mbox{\small spin}}$ is the percentage spin shift relative to the Larmor frequency of the
$^{17}$O nucleus and is attributable only to the electron spin susceptibility. The normal state value is determined independently
from the extrapolation to high field of the  difference \mbox{$^{17}\!K(T = 100$ K$)$} - \mbox{$^{17}\!K(T = 20$ K$)$}. This
expression extrapolates to the normal state
$^{17}\!K_{\mbox{\small spin}}$ because other contributions to the total spectrum shift are either negligible at high fields
($\Delta H_{\mbox{\small dia}}$ and $\Delta H_{\mbox{\small v}}$), temperature independent ($^{17}\!K_{\mbox{\small orb}}$),
or both (quadrupolar terms).  The decrease of
$^{17}\!K_{\mbox{\small spin}}$ with decreasing 
\begin{figure}[h]
\centerline{\epsfxsize0.90\hsize\epsffile{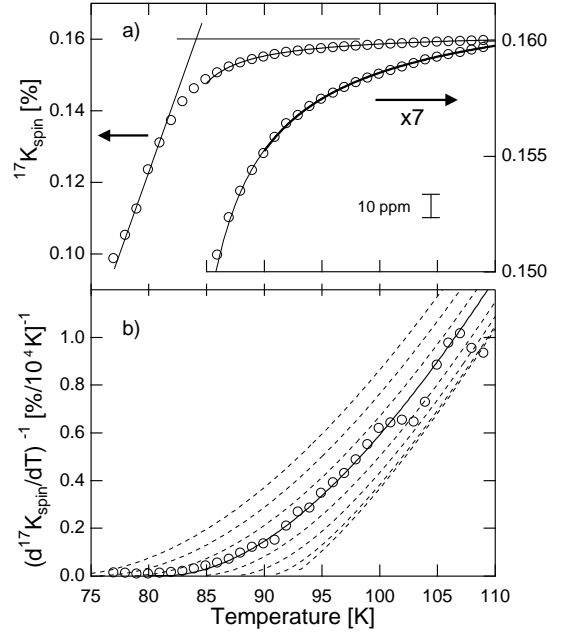}}
\begin{minipage}{0.95\hsize}
\caption[]{\label{KvT84T}\small
   $^{17}$O(2,3) spin shift at 8.4 T for $H || {\hat c}.$ The theoretical calculation
taking into account 2D pairing fluctuations is shown as curves.  a) The inset shows a factor of 7 
expanded vertical scale demonstrating excellent agreement between theory and 
experiment. The crossed lines determine $T_c$, as discussed in the text.
b) The inverse of the derivative, $(d\,^{17}\!K_{\mbox{\small spin}}/d T)^{-1}$, as
discussed in the text.
The dashed curves indicate calculations for $T_{cmf}=72$ K, 75 K, 78 K, 80.9 K,
84 K, 87 K, 90 K, 92.5 K.
}
\end{minipage}
\end{figure}
\noindent 
temperature is smooth, showing no 
   discontinuities in either 
magnitude or slope, suggestive of a crossover region. The 
crossover 
   shifts to lower temperatures as the field is increased.

Peak frequencies were determined by nonlinear 
   least-squares fits to a gaussian around the peak region. In our stable superconducting magnet (8.4 T) this method 
allowed us to achieve precision in determination of the 
peak frequency of 1 part per million, corresponding to better than one part in a thousand precision relative to the total normal state
value of
$^{17}\!K_{\mbox{\small spin}}$.
For data obtained with the 
Bitter magnet, corrections are required for variations of magnetic field with cooling water temperature, and thus the
precision of the peak frequencies is only 5 parts per million.    In
\mbox{Fig.  \ref{KvT84T}a} we plot $^{17}\!K_{\mbox{\small spin}}$ at
\mbox{8.4 T} with an expanded version in the inset but 
   on the same temperature scale. The $^{17}\!K_{\mbox{\small spin}}$ data have a monotonic decrease easily discernible 
   below 110 K. 

   The precision of our spin shift measurements at 8.4 T allows a quantitative comparison with the theory of 
   superconducting fluctuations.  
   Pairing fluctuation corrections to the 
Pauli spin susceptibility 
in the zero field limit have been 
considered theoretically by several
authors
.\cite{randeria94,fulde70}
Gaussian 
fluctuation corrections diverge 
at the mean-field transition 
temperature $T_{cmf}$.
\begin{figure}
\centerline{\epsfxsize0.90\hsize\epsffile{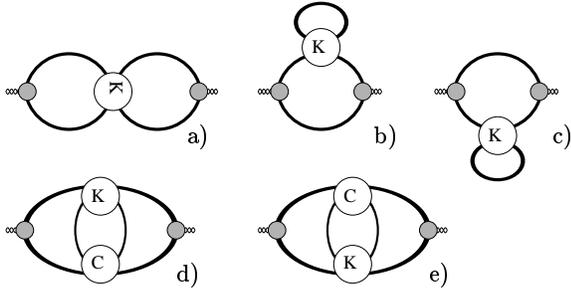}}
\begin{minipage}{0.95\hsize}
\caption[]{
\label{diagrams}
Feynman diagrams for the pairing fluctuation corrections to the Pauli spin susceptibility, to leading order in $T_{cmf}/E_F$, as discussed in the text.
}
\end{minipage}
\end{figure}
\noindent 
Calculations for
fluctuation contributions to the spin susceptibility 
based on long wavelength static fluctuations at zero field near $T_{cmf}$
predict 
$(d \delta\chi_{s}/d T)^{-1} \propto T-T_{cmf}$ in 2D,\cite{randeria94}
which cannot explain the curvature in our measurements shown
in \mbox{Fig. \ref{KvT84T}b}, and
$(d \delta\chi_{s}/d T)^{-1}\propto \sqrt{T-T_{cmf}}$ in 3D,\cite{fulde70} producing curvature opposite to that of our measurements.

We calculate $\chi_s$ in
the weak-coupling limit for a quasi-2D, d-wave superconductor taking into account Landau quantization of the orbital motion of
pairs by a magnetic field ($H||\hat{c}$).
Our calculations include dynamical pairing fluctuations
and we sum over all Landau levels in order
to extend the range of validity of the theory
to higher fields and temperatures. 
A detailed presentation of the theory is given in 
Eschrig {\it et al.};\cite{Eschrig98}
here we provide a short summary.
The pair fluctuation propagator for $d$-wave pairing with coupling constant $g$
is given by $ L(Q)^{-1} = g^{-1}-T\sum_{\epsilon_n} B_2(\epsilon_n,Q)$,
where $B_2(\epsilon_n,Q)$ is an impurity renormalized two-particle
susceptibility in the $d$-wave pairing channel as discussed 
in Mitrovi{\'c} {\it et al.}\cite{mitrovic98}
and in Eschrig {\it et al.}\cite{Eschrig98}
We use the notation $Q=(\omega_l,\vec{q} )$, which combines Matsubara energy, $\omega_l=2\pi lT$, and
pair momentum, $\vec{q}$, of the fluctuation mode; the latter
is quantized in a magnetic field.
The results we obtain derive from the summation of all 
leading order pairing fluctuation corrections in $T_{cmf}/E_F$,
($E_F$ is the Fermi energy renormalized by exchange field corrections),
given by the diagrams
shown in \mbox{Fig. \ref{diagrams}} (a-e), the Maki-Thompson (a), the
density-of-states (b,c), and the Aslamazov-Larkin (d,e) contributions.
$K$ denotes the fluctuation propagator
renormalized by external vertex corrections
due to impurities. In contrast to the fluctuation corrections to the
spin-lattice relaxation rate the contributions
(d,e) in \mbox{Fig. \ref{diagrams}} have the same order in 
$T_{cmf}/E_F$ as (a-c).
However, they contain only one singlet pair fluctuation mode, $K$,
the other mode in the particle-particle channel is a triplet
impurity Cooperon-like mode, $C$.
In the clean limit diagrams (d,e) vanish, in the dirty limit they give
the main contribution. For intermediate impurity scattering
all diagrams contribute significantly.
It is possible to write the sum of all diagrams in \mbox{Fig. \ref{diagrams}}
in a compact way.  
Considering $B_2(\epsilon_n,Q)$ as a functional of the quasiparticle 
impurity self energy $\Sigma (\epsilon_n)$,
the sum of all leading order pairing fluctuation corrections 
to $\chi_s$ can be written as,\cite{Eschrig98}
\begin{eqnarray}
\label{chi}
\delta \chi_s&=&  (^{17}\gamma \hbar)^2 \sum_{n,Q}
\frac{\delta^2 B_2(\epsilon_n,Q)}{\delta \Sigma(\epsilon_n)^2}L(Q).
\end{eqnarray}

In \mbox{Fig. \ref{KvT84T}} our calculation for 8.4 T is compared with experiment.  The parameters
extracted from the fit yield
$E_F= 930 \pm 30$ meV, and $T_{cmf} = 80.9 \pm 0.3$ K. 
We used the same scattering parameters 
as in our comparison of the theory of dynamical pairing fluctuations
with the field dependence of spin-lattice relaxation in
Mitrovi{\'c} {\it et al.}\cite{mitrovic98}
Dynamical fluctuations and orbital quantization effects produce the curvature shown 
in \mbox{Fig. \ref{KvT84T}b}.  
The fit to our theory is performed 
in the region $T > 90$ K, and is shown by the heavy 
solid curve in \mbox{Fig. \ref{KvT84T}a}. Extension of the same fit to lower 
   temperatures, as indicated by the thin solid curve, demonstrates that 
the theory fits the data well down to $T = 85$ K. 
Below this temperature critical fluctuations become significant.

\begin{figure}[h]
\centerline{\epsfxsize1.0\hsize\epsffile{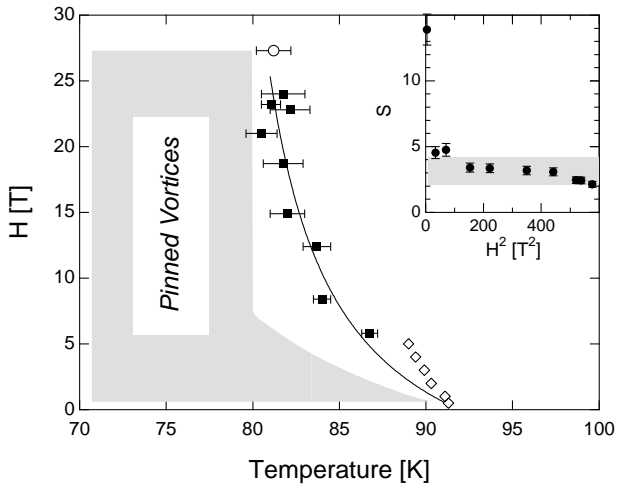}}
\begin{minipage}{0.95\hsize}
\caption[]{\label{HvT}\small
   $H-T$ phase diagram from electron spin susceptibility as determined by $^{17}$O NMR for YBCO$_{6.95}$. The dark 
   squares represent a determination of $H_{c2}(T)$ as discussed in the text. 
The curve is a guide to the eye.
 Vortices become pinned in the cross-hatched 
   region.\cite{reyes97a} The open circle indicates onset of vortex pinning, an extension of the earlier work 
   of Bachman {\it et al.}\cite{reyes97a} Open diamonds are from diamagnetism measurements.\cite{welp89}
   Inset: Magnetic field dependence of ($S = [d(^{17}\!K_{\mbox{\small spin}})/dT]_{max} \times [T_c/^{17}\!K_{\mbox{\small spin}}(120 K)]$).

}
\end{minipage}
\end{figure}

   From the temperature dependence of the spin susceptibility in the mean field regime of the superconducting state it 
is possible to 
   extrapolate linearly back to the susceptibility of the normal state to determine a crossover temperature, $T_c$. We determined the linear
temperature dependence of the  NMR spin shift in the superconducting state from the  maximum slope, 
$d(^{17}\!K_{\mbox{\small spin}})/dT$, and performed an extrapolation to the normal state shift of 0.16$\, $\% to find $T_c$; see \mbox{Fig.
\ref{KvT84T}a}. The results are presented in
\mbox{Fig.
\ref{HvT}}. In particular, at 8.4 T, we found $T_c = 84$ K $\pm \, 0.5$ K. The
slopes we used are shown in the inset of \mbox{Fig. \ref{HvT}} for various fields.  Because the slope is expected to scale
with $\Delta^2$,
   we can confirm the validity of our approach by comparing the 
slopes in \mbox{Fig. \ref{HvT}} with 
direct measurements of the energy gap, $\Delta$.  Tunneling and photoemission 
   measurements \cite{schabel97} suggest a gap of $2\Delta /k_B T_c = 6 \, \pm \, 1$ for YBCO. Using the mean field result for $\chi_s$ without
exchange corrections, but 
allowing for gap anisotropy, we estimate the scaled slope for d-wave pairing to be in the range  2.1 to 4.2, shown
in the inset of
\mbox{Fig. \ref{HvT}} as a cross-hatched region. The anomalously large slope 
   at 2.1 T can be attributed to 
diamagnetic  contributions to the frequency  shift measurement. 
For all other fields the experimental slopes presented in the inset of 
\mbox{Fig. \ref{HvT}} are consistent with our expectation for the slope of $\chi_s$.
The crossover line, $H_{c2}(T)$, in \mbox{Fig. \ref{HvT}} 
exhibits upward curvature. Qualitatively similar behavior, but at lower
fields, has been reported  in specific heat experiments.\cite{roulin96}
	With decreasing temperature there is a smooth crossover from the fluctuation
regime to a vortex liquid phase.
Deviations of our theory from experiment 
in \mbox{Fig. \ref{KvT84T}}
below 85 K indicate onset of critical 
fluctuations.
The smooth crossover suggests a relationship
between critical fluctuations and the 
vortex liquid phase.
\noindent
However, there is no adequate theory for the interplay between 
critical fluctuations and the fluctuating currents in the vortex liquid.

The vortex liquid phase becomes well established at temperatures less than those given by $H_{c2}(T)$. At lower temperatures vortices become
pinned, as we have shown previously.\cite{reyes97a} Using
two 
   independent methods we determined a region of the phase diagram in which pinned vortices are 
   present, which we show as the hatched region of \mbox{Fig. \ref{HvT}}.
We have extended that work to 27.3 T, shown as an open
circle in \mbox{Fig. \ref{HvT}}, where the onset of vortex pinning is observed by spin-spin relaxation.
For low fields we point out that the pinning temperature and melting
temperature\cite{kwok94,roulin96}
   of untwinned, single crystals coincide. Our understanding of vortex pinning in combination with our new measurements of $H_{c2}(T)$ show
that the region of liquid vortex matter is restricted to 
   both high temperatures and low magnetic fields. 

   In summary, we find that $^{17}$O NMR Knight shifts give a precise determination of the temperature dependence of 
   the Pauli spin susceptibility, $\chi_s$.  We find significant rounding near $T_c$ indicating that superconducting 
   fluctuations smear the transition. Consequently, the transition is best represented as a crossover 
   from normal state behavior to that of a vortex liquid.  Our calculations for superconducting fluctuations, taking into 
   account dynamical pairing fluctuations and the effects of orbital quantization, are in excellent agreement with
experiment in the temperature range down to $T_c$. The decreasing susceptibility with decreasing temperature above
$T_c$ can be fully accounted for by the opening of a pairing pseudogap. Finally, we have established the
$H_{c2}(T)$ phase diagram for YBa$_2$Cu$_3$O$_{6.95}$ up to 24 T.

           We gratefully acknowledge useful discussions with G. Crabtree,
M. Fogelstr\"om, K. Poeppelmeier, D. Rainer, 
   H. Safar, Y. Song, and S.-K. Yip.  We are particularly thankful to C. Hammel for providing the sample.  This 
   work is supported by the National Science Foundation (DMR 91-20000) through the Science and Technology 
   Center for Superconductivity.  The work at the National High Magnetic Field Laboratory was supported by the 
   National Science Foundation under Cooperative Agreement No. DMR95-27035 and the State of Florida.
   ME acknowledges support from the Deutsche Forschungsgemeinschaft.
\vspace{-0.5cm}
\bibliographystyle{unsrt}

\vspace{3cm}
\end{multicols}
\end{document}